\documentclass[reprint, amsmath,amssymb, aps,nofootinbib]{revtex4-1}
\pdfoutput=1
\usepackage{graphicx,subfigure,color}
 \usepackage{dcolumn}
\usepackage{bm}
\usepackage{hyperref}
\usepackage[mathlines]{lineno} %

\DeclareMathAlphabet{\mathpzc}{OT1}{pzc}{m}{it}
\let\a=\alpha \let\b=\beta \let\g=\gamma \let\d=\delta \let\e=\epsilon
  \let\th=\theta  \let\k=\kappa
\let\l=\lambda \let\m=\mu \let\n=\nu \let\x=\xi \let\p=\pi 
\let\s=\sigma   \let\f=\phi  
 
        \let\Th=\Theta 
\let\X=\Xi  \let\S=\Sigma  \let\Y=\Psi
 
\let\la=\label  
  
 \def\bd{\begin{document}} \def\ed{\end{document}}
\def\ds{\documentstyle} \let\fr=\frac \let\bl=\bigl \let\br=\bigr
\let\Br=\Bigr \let\Bl=\Bigl
\let\bm=\bibitem
\let\na=\nabla
\def\tU{{\widetilde U}}
\let\pa=\partial \let\ov=\overline
\def\ie{{\it i.e.\ }}
\newcommand{\be}{\begin{equation}}
\newcommand{\ee}{\end{equation}}
\def\ba{\begin{array}}
\def\ea{\end{array}}
\def\ft#1#2{{\textstyle{{\scriptstyle #1}\over {\scriptstyle #2}}}}
\def\fft#1#2{{#1 \over #2}}
\def\F#1#2{{ F_{#1}^{(#2)} }}
\def\cF#1#2{{ {\cal F}_{#1}^{(#2)} }}
 \def\bchi{{\boldsymbol \chi}}
  \def\bzeta{{\boldsymbol \zeta}}
    \def\bw{{\boldsymbol \omega}}
    \def\bfe{{\boldsymbol \epsilon}}
\def\R{{\bf R}}
\def\sst#1{{\scriptscriptstyle #1}}
\def\oneone{\rlap 1\mkern4mu{\rm l}}
\def\e7{E_{7(+7)}}
\def\td{\tilde}
\def\wtd{\widetilde}
\def\im{{\rm i}}
\def\bog{Bogomol'nyi\ }
\newcommand{\ho}[1]{$\, ^{#1}$}
\newcommand{\hoch}[1]{$\, ^{#1}$}
\newcommand{\bea}{\begin{eqnarray}}
\newcommand{\eea}{\end{eqnarray}}
\newcommand{\ra}{\rightarrow}
\newcommand{\lra}{\longrightarrow}
\newcommand{\Lra}{\Leftrightarrow}
\newcommand{\cB }{{\cal B}}
\newcommand{\cO}{{\cal O}}
\newcommand{\vecx}{\vec{x}}
\newcommand{\vecy}{\vec{y}}
\newcommand{\vecp}{\vec{p}}
\newcommand{\vecq}{\vec{q}}
\newcommand{\tr}{{\rm tr} }
\newcommand{\Tr}{{\rm Tr} }
\newcommand{\NP}{Nucl. Phys. } 
\newcommand{\cL}{{\cal L}}
\newcommand{\cA}{{\cal A}}
\newcommand{\cT}{{\cal T}}
\newcommand{\cD}{{\cal D}}
\newcommand{\cH}{{\cal H}}
\def\vev#1{\langle{#1}\rangle}

\def\sst#1{{\scriptscriptstyle #1}}
\def\0{{\sst{(0)}}}
\def\1{{\sst{(1)}}}
\def\2{{\sst{(2)}}}
\def\3{{\sst{(3)}}}
\def\4{{\sst{(4)}}}
\def\5{{\sst{(5)}}}
\def\6{{\sst{(6)}}}
\def\7{{\sst{(7)}}}
\def\8{{\sst{(8)}}}
\def\9{{\sst{(9)}}}
\def\p{{\sst{(p)}}}
\def\q{{\sst{(q)}}}
\def\ve{\varepsilon}
\def\vf{\varphi}
\def\F{\Phi}
\def\wg{\wedge}

\def\thb{\bar{\theta}}
\def\Thb{\bar{\Theta}}
\def\barp{\bar{p}}
\def\barq{\bar{q}}
\def\barc{\bar{c}}
\def\bard{\bar{d}}
\def\e{\epsilon}

\def \bi{\bibitem}
\def \la {\label}

\def \l {\lambda}
\def\foot{\footnote}
\def \tl  {{\tilde \l}}
\def \sql {{\sqrt \l}}
\def \adss {$AdS_5 \times S^5$\ }
\newcommand{\rf}[1]{(\ref{#1})}
\def \ov {\over}
\def\dla{{\tiny \frac{d}{d\lambda}}}
\def \dep{\frac{d}{d\varepsilon}}

\def\th{\theta}
\def\Th{\Theta}
\def\vth{\vartheta}
\def\btheta{{\bar\theta}}
\def\ttheta{{{\tilde\theta}}}
\def\bttheta{{{\bar\ttheta}}}
\def\vth{\vartheta}
\def\ve{\varepsilon}
\def\ra{\rightarrow}
\def\N{\nabla}
\def\F{{\cal F}}
\def\uM{\underline{M}}
\def\uA{\underline{A}}
\def\uN{\underline{N}}
\def\uP{\underline{P}}
\def\ua{\underline{a}}
\def\ub{\underline{b}}
\def\uc{\underline{c}}
\def\ud{\underline{d}}
\def\ue{\underline{e}}
\def\uf{\underline{f}}
\def\ui{\underline{i}}
\def\uj{\underline{j}}
\def\uk{\underline{k}}
\def\ul{\underline{l}}
\def\ual{\underline{\alpha}}
\def\ube{\underline{\beta}}
\def\um{\underline{m}}
\def\un{\underline{n}}
\def\up{\underline{p}}
\def\uq{\underline{q}}
\def\ur{\underline{r}}
\def\us{\underline{s}}
\def\umu{\underline{\mu}}
\def\unu{\underline{\nu}}
\def\ula{\underline{\l}}
\def\uka{\underline{\k}}
\def\usi{\underline{\s}}
\def\urh{\underline{\r}}
\def\cc{\circ}
\def\eqv{\equiv}

\def\ni{\noindent}

\def\Ep{E^{{}^{(+)}}}
\def\Em{E^{{}^{(-)}}}

\def\Mp{M^{{}^{(+)}}}
\def\Mm{M^{{}^{(-)}}}

\def \ha{{1\ov 2}}

\def\r{\rho}
\def\Y{{\rm Y}}
\def\X{{\rm X}}
\def\tY{\tilde{\rm Y}}
\def\tX{\tilde{\rm X}}
\def\dY{\dot{\rm Y}}
\def\dX{\dot{\rm X}}

\def \J {\mathcal{J}}
\def \del {\partial}

\def\dF{\dot{F}}
\def\dG{\dot{G}}
\def\df{\dot{f}}
\def \E {{\cal E}}
\def \S {{\cal S}}
\def \J {{\cal J}}

\def\ms{\mathcal{S}}
\def\mj{\mathcal{J}}
\def\soj{\fr{\ms}{\mj}}
\def \R {{\bf R}}
\def \om {\omega}
\def \bE {\bar E}
\def \x {{\cal X}}

\def \bi{\bibitem}
\def \la {\label}

\def \l {\lambda}
\def\foot{\footnote}
\def \tl  {{\tilde \l}}
\def \sql {{\sqrt \l}}
\def \adss {$AdS_5 \times S^5$\ }
\def \ov {\over}

\def \varpi {{\rm w}}

\def\thb{\bar{\theta}}
\def\Thb{\bar{\Theta}}
\def\mb{\bar{\m}}
\def\ab{\bar{\a}}
\def\zb{\bar{z}}
\def\psib{\bar{\psi}}
\def\barp{\bar{p}}
\def\barq{\bar{q}}
\def\barc{\bar{c}}
\def\bard{\bar{d}}
\def\e{\epsilon}
\def\wb{\bar{w}}
\def\lb{\bar{\l}}
\def\Jb{\bar{J}}
\def\Nb{\bar{N}}
\def\Zb{\bar{Z}}
\def\pab{\bar{\pa}} 

\def\At{\tilde{A}}
\def\Bt{\tilde{B}}
\def\Ct{\tilde{C}}
\def\Dt{\tilde{D}}
\def\Et{\tilde{E}}
\def\Ft{\tilde{F}}
\def\Gt{\tilde{G}}
\def\Ht{\tilde{H}}
\def\Kt{\tilde{K}}
\def\Mt{\tilde{M}}
\def\Nt{\tilde{N}}
\def\Rt{\tilde{R}}
\def\at{\tilde{a}}
\def\bt{\tilde{b}}
\def\ct{\tilde{c}}
\def\dt{\tilde{d}}
\def\et{\tilde{e}}
\def\ft{\tilde{f}}
\def\htil{\tilde{h}}
\def\gt{\tilde{g}}
\def\nt{\tilde{n}}
\def\mut{\tilde{\mu}}
\def\nut{\tilde{\nu}}
\def\pht{\tilde{\f}}
\def\vft{\tilde{\vf}}

\def\rht{\tilde{\rho}}

\def\asth{\hat{*}}
\def\phh{\hat{\phi}}

\def\bA{{\bf A}} 
\def\ola{\overleftarrow}
\def\ora{\overrightarrow}
\def\alt{\tilde{\a}}

\def\eh{\hat{e}}
\def\eph{\hat{\e}}
\def\ph{\hat{p}}
\def\alh{\hat{\a}}
\def\beh{\hat{\b}}
\def\gah{\hat{\g}}
\def\Fh{\hat{F}}
\def\muh{\hat{\m}}
\def\nuh{\hat{\n}}
\def\thh{\hat{\th}}
\def\rhh{\hat{\r}}
\def\dh{\hat{d}}
\def\ih{\hat{i}}
\def\jh{\hat{j}}
\def\hh{\hat{h}}
\def\nh{\hat{n}}
\def\gh{\hat{g}}
\def\kh{\hat{k}}
\def\deh{\hat{\d}}
\def\wh{\hat{w}}
\def\lah{\hat{\l}}
\def\Ah{\hat{A}}
\def\Kh{\hat{K}}
\def\Nh{\hat{N}}
\def\Rh{\hat{R}}
\def\Ch{\hat{C}}
\def\Omh{\hat{\Omega}}

\def\xh{\hat{x}}

\def\ps{\rlap{\, /}\;\,p }
\def\ks{\rlap{\, /}\;\,k }

\def\gym{g_{YM}}

\def\adot{\dot{a}}
\def\bdot{\dot{b}}
\def\bpa{\bar{\pa}}

\def\pr{\prime}
\def\ssk{\medskip}
\def\clb{\color{blue}}
\def\clr{\color{red}}
\def\clg{\color{green}}

\def\bfA{{\bf A}}
\def\bfB{{\bf B}}
\def\bfK{{\bf K}}
\def\bfU{{\bf U}}
\def\bfX{{\bf X}}
\def\bfY{{\bf Y}}
\def\bfZ{{\bf Z}}
\def\bfg{{\bf g}}
\def\bfn{{\bf n}}

\def\bfL{{\bf L}}
\def\bfJ{{\bf J}}
\def\bfE{{\bf E}}
\def\bfF{{\bf F}}
\def\bfQ{{\bf Q}}
\def\bfC{{\bf C}}

\begin{document}
\title{Complete Einstein equation from the  generalized First Law of  Entanglement } 
\author{Eunseok Oh$^{1}$, I.Y. Park$^{2}$ and Sang-Jin Sin$^{1}$
}
 \email{lspk.lpg@gmail.com, inyongpark05@gmail.com, sangjin.sin@gmail.com}
 \affiliation{
$^{1}$
Department of Physics, Hanyang University, Seoul 04763, Korea.\\
$^{2}$ \it Department of Applied Mathematics,
Philander Smith College  Little Rock, AR 72223, USA 
	}  \date{\today}%
\begin{abstract}
Recently it was observed that the  first law of Entanglement   leads to the linearized Einstein equation.  
In this paper, we point out that  the gravity dual of an relative entropy expression is equivalent  to the full non-linear Einstein equation. We also construct an entanglement vector field $V_{E}$ whose flux is the entanglement entropy. The flow of the vector field  looks like sewing two space regions  along the interface.  
\begin{description} 
\item[PACS numbers]11.25.Tq, 03.65.Ud, 04.25.dg
\end{description}
\end{abstract}
\pacs{here}
\keywords{ Entanglement Entropy, Einstein Equation} 
\maketitle

\section{Introduction}
One of most inspiring ideas in recent development of string theory is the  suggestion  \cite{VanRaamsdonk:2010pw,Maldacena:2013xja}  that  the classical spacetime is  a consequence of the quantum entanglement without which   two nearby  regions of spacetime would take apart  \cite{VanRaamsdonk:2010pw,Maldacena:2013xja} and  moreover, the Einstein equation   itself is coming from  a relation  of entanglement entropy at least in linearized level \cite{Lashkari:2013koa}.  The latter is a consequence of connecting  two different  descriptions of entanglement entropy (EE): one  as the  area of Ryu-Takayanagi  surface  \cite{Ryu:2006ef} and the other 
 as the expectation value of the modular Hamiltonian \cite{Casini:2011kv}.  
Later,  it was pointed out  \cite{Faulkner:2013ica} that 
such relation between the first law of EE and linearized gravity equation are connected through the off-shell Noether theorem   formulated by  Wald \cite{Wald:1993nt,Iyer:1994ys,Iyer:1995kg,Wald:1999wa,Hollands:2012sf}.  

 Deriving the Einstein equation from the first law has much similarity to the activity of 90's lead by  
 the work of Jacobson\cite{Jacobson:1995ab}: assuming the thermodynamic first law he got  the gravity equation. 
The difference of the recent activity \cite{Lashkari:2013koa,Faulkner:2013ica}  is that  the entanglement first law and its gravity dual  themselves are derived from the conformal field theory (CFT) although it gave only linearized equation.  
That is,  recent activities  aim  to derive the  Einstein equation of  the dual gravity  of a CFT assuming the presence of holography.
In ref. \cite{Faulkner:2017tkh}, the authors  extended the program to the non-linear   second order in perturbative scheme. The major efforts of  ref.  \cite{Faulkner:2017tkh} is devoted to 
derive  the 'gravity dual expression of the relative entropy' (GDERE) starting from CFT up to second order. 

While   proving the GDERE from the CFT to all order is yet to be done, we can still ask   
"if we assume this part is done,   does  it imply the full non-linear Einstein equation?" 
The goal of this paper is to prove that the answer is yes. 
As we will see later,  having the  GDERE  gives the 'gravitational form of  generalized  first law of   entanglement entropy' and it is equivalent to the Einstein equation.

The other goal of this paper is to  construct a vector field associated with the EE whose flux is the EE independent of the surface  over which the vector field  is integrated. 
The flux line, once the total flux  quantized,  is analogous to the microscopic  worm-hole and concentrated along the boundary of the entangled regions.  

\vskip 0.2cm
\section{Einstein Equation from Entanglement in linear order }
To set up notation, we start  with a short review of relevant concepts. 
Given a physical state given by a density matrix $\rho$ and a ball like region $B$ of radius $R$, one can 
decompose the Hilbert space into tensor product ${\cal H}={\cal H}_{B}\otimes {\cal H}_{\bar B}$, where 
${\cal H}_{B}$ is the Hilbert space of local fields over $B$. 
The reduced density operator $\rho_{B} =\Tr_{{\cal H}_{\bar B}}\rho$. The entanglement entropy
is given by $S_{B}=-\Tr \rho_{B} \ln \rho_{B}$. 
From now on, we delete the subscript   $B$ when there is no confusion. 
The modular Hamiltonian  $H_{0}=-\log \rho_{0}$ for a reference state $\rho_{0}$ which is normalized by $\Tr\rho_{0}=1$. 
If we call the expectation value  of the modular Hamiltonian for the state $\rho$  as the `Energy' of the state $\rho$,  
then we have 
$
E=\vev{H_{0}}= -\Tr \rho \ln \rho_{0}
$. 
Under  finite  variation  of the state  from $\rho_{0}$ to $\rho$,  we have following  identity 
\bea 
\Delta E- \Delta S &=& S(\rho|\rho_{0}),  \label{cft1}\\
 \hbox{ where}\quad   \Delta E &=& -\Tr (\rho-\rho_0)  \ln \rho_{0} ,  \label{cft2} \\
 \Delta   {S} &=& -\Tr \rho \ln \rho +  \Tr \rho_{0} \ln \rho_{0}, \label{cft3} \\
 S(\rho|\rho_{0}) &=& \Tr\rho(\ln\rho-\ln\rho_{0})).   \label{cft4}   
  \eea 
Three  important remarks are in order. First,   (\ref{cft1}) and  (\ref{cft4}) can be used as 
 the definition and a  result interchangeably. Second, $\Delta E$ is not a total variation
while $\Delta S$ is, because  the relative entropy,  $S(\rho|\rho_{0})$,  can not be so. 
Similar phenomena will be observed in their gravitational versions. 
 Finally, the relative entropy is  alway positive \cite{Blanco:2013joa} and 
  this is the origin of the   entanglement first law: as a function of $\rho$, 
   $S(\rho|\rho_{0})$  is  minimal at the reference state. Such  extremality  condition 
   is   the usual entanglement first law,
 \be
  \delta E-\delta S=0,  \label{1st}\ee
  where $\delta f= \dla f|_{ \lambda=0}$ for   $f$  which is a  one parameter family of $\lambda \in [0,\ve]$. 
The  positivity of the relative entropy is also related to  
the positivity of energy \cite{Lin:2014hva,Lashkari:2016idm} and that of Fisher metric for information theory \cite{Lashkari:2015hha}.  
Both  terms   of the first law  can be calculated in gravitational languages using the AdS/CFT and Ryu-Takayanagi formula and  it  turns out that  the first law leads to the linearized Einstein equation as we will review below.

Suppose the density opereator depends on parameters $R^1,R^2,\cdots,R^M$ which we symbolically denote by a vector $\bf R$ and let  $\rho_0=\rho({\bf R}_0)$ and $\rho=\rho({\bf R}_1)$ for some ${\bf R}_0,{\bf R}_1$.  
 Introducing  the modular potential $V=-\ln \rho$ and the modular force 
 $F_{\alpha}=-\nabla_\alpha V  $
 in the parameter space,  we can express the relative entropy as 
\be
S(\rho|\rho_{0})= \vev{\int_C d{\bf R} \cdot {\bf F}    }   , 
\ee
which can be interpreted 
 as the `work', $W$, done  {\it on} the system by   $ {\bf F} $ 
  to change  the system from $\rho_0$ to $\rho$. 
  Notice that it is independent of the path C connecting $\rho_0$ and $\rho$ of the integration.  
  Then the identity  (\ref{cft1}) itself, although  in a finite difference form,  can be considered  as  a 
 first law', 
 \be
\Delta E- \Delta S=W =S(\rho|\rho_{0}) \label{gen1st} . 
\ee
which we call    `generalized entanglement first law'. 
In fact, it has a gravity version.  
Our claim is that while we get the linearized gravity equation by using  (\ref{1st}), we will get the full non-linear equation if we use the gravity version of   (\ref{gen1st}).  

For any CFT vacuum  $\rho_{0}=|0\rangle\langle 0|$, a confomal mapping can be constructed which maps the causal development of the ball $B$ to  a hyperbolic cylinder $H^{d-1}\times R_{\tau}$
  and  $\rho_{0}$  to  a thermal density  operator $\exp(-2\pi RH_{\tau})$  of CFT on hyperbolic space.  Namely, the vacuum state is mapped to  a thermal state of   temperature $T =1/2\pi R$  on the $H^{d-1}$ and modular Hamiltonian actually generates the time evolution of CFT on the hyperbolic space. 
According to the AdS/CFT the thermal state on $H^{d-1}$  can be represented by a AdS black hole with temperature $T =1/2\pi R$, the AdS-Rindler space, which can be figured as a patch of AdS 
space with Poincare metric. 

As described above, the Hamiltonian $H_{\tau}=\int_{H^{d-1}}T_{tt}$   is  equal to the Unitarily  transformed  modular hamiltonian of the original CFT in the flat space  \cite{Casini:2011kv}: $H_{0}=2\pi R U {\tilde H_{\tau}}U^{-1}$. 
Using this, the authors of \cite{Casini:2011kv} expressed the modular Hamiltonian 
$H_{0}$ in terms of energy momentum tensor of CFT  
\be
H_{0} =2\pi \int_{B}d^{d-1}x \fr{R^2-|\vec{x} |^2}{2R}T_{tt} \la{mH}
	= \int_{B}d\sigma^{\mu} \zeta_{B}^{\nu} T_{\mu\nu} , 
\ee
where $\vec{x}=0$ is located at the center of the ball of radius $R$ and 
 $\zeta_{B}^{\mu}$ is the pullback of the killing vector $\frac{\partial}{\partial{\tau}}$ by the mapping that maps the causal development of $B$ to the hyperbolic cylinder $H^{d-1}\times R_{\tau}$. 
It can be considered as the boundary restriction of a Killing vector $\xi$ of AdS which vanishes at $\tilde B$. 
More explicitly 
\bea
\xi_{B}=
\frac {\pi}{R} [R^{2}-z^{2}-t^{2}-x^{i}x_{i}]\partial_{t}-\frac {2\pi}{R} t[z\partial_{z}+x^{i}\partial_{i}] ,  \label{killing}
\eea
and 
$\zeta_{B}=\lim_{z\to 0} \xi_{B} $. The entanglement energy $E_{B}$ is given by $E_{B}= \int_{B} \zeta_{B}^{\mu} \vev{T_{\mu\nu}}d\sigma^{\nu}$. 
 Now, the gravitational dual of $\delta E_{B}$ is readily given since AdS/CFT dictionary gives the relation
between the expectation value of energy momentum tensor and the metric variation, 
 $\vev{T_{\mu\nu}} \sim  z^{d-2}\delta g_{\mu\nu}$. 
The gravitational dual of   $\delta S_{B}$ can be   given using the  Ryu-Takayanagi prescription 
$\S_{B}={\rm Area}[\tilde B]/4G_{N}$ \cite{Ryu:2006ef}.
\def\bchi{{\boldsymbol \chi}}
\def\beps{{\boldsymbol \epsilon}}
The crucial observation of \cite{Faulkner:2013ica} is that there exists a $d-1$ form $\bf \chi$  in asymptotic $AdS_{d+1}$ such that 
\be 
\int _{B} \bchi = \delta E_{B}^{grav} , \quad {\rm and}\quad \int _{\tilde B} \bchi = \delta S_{B}^{grav}
\ee 
based on the formalism of Iyer-Wald\cite{Iyer:1994ys,Iyer:1995kg}:
\be 
\delta E_{B}^{grav}-\delta S_{B}^{grav}= \int _{B-{\tilde B}} \bchi =   \int _{\Sigma} d\bchi,  \label{EE1st}
 \ee 
where   $\Sigma$ is $t=0$ slice whose boundaries are $B$ and $\tilde B$. 
Since it turnes out to be 
\be 
d\bchi=-2\xi^{a}_{B} \delta E_{ab}{\beps}^{b}, \label{linearE}
\ee 
the entanglement first law implies the linearised Einstein equation $ \delta E_{ab}=0$. 

 Since understanding    Wald's  formalism is essential for later formalism, 
 we describe it below shortly. 
Start from the Lagrangian written in differentiable form notation: 
$\bfL\equiv  L[\phi] {\boldsymbol \e} $,
where $\f$ is a collective representation of the bulk fields including the   metric and ${\boldsymbol \e}$ is the volume form.
The general  variation of $\bfL$ can be written as
\bea
\d \bfL [\phi]=\bfE^{\phi} \d\f+d\Th[\d \phi] , \label{varL} 
\eea
 where $\bfE^{\phi}$ denotes   field equations and $\Th$ the symplectic potential current that contains Gibbons-Hawking term. When the variation is a 
 diffeomorphism   generated by a vector field $\xi$,  $\delta_{\xi}  \bfL  =d( \xi\cdot \bfL )$ since  
 $\delta_{\xi}=i_{\xi}d +d i_{\xi}$ and $\bfL$ is the top form. 
 In terms of  the Noether current codimension 1 form 
  \bea
\bfJ_\xi=\Th[\d_\xi \f]-\xi\cdot \bfL  \la{J1} , 
\eea 
Eq. (\ref{varL})   for the diffeomorphic variation is  
 \be
 d\bfJ_\xi= - \bfE^\phi\cdot \d_\xi\f,
 \ee
  so that $\bfJ$ is the closed form for the fields at on-shell. Therefore $\bfJ_\xi=d\bfQ_\xi$ at  on-shell. For  off-shell,  one can show \cite{Iyer:1995kg, Faulkner:2013ica}  that  
\bea
\bfJ_\xi=d\bfQ_\xi + \xi^{a}\bfC_{a}, \label{J2}
\eea
where $\bfC_{a}$'s are constraints which vanish for metric satisfying the  equation of motion
\cite{Iyer:1994ys}:
\bea 
\bfQ &=& \frac{1}{16\pi G_{N}} \nabla^{a}\xi^{b} \beps_{ab}, \quad  
\bfC_{a}=2E^{g}_{ab}\beps^{b}, \label{constraint}\\
 \hbox{with } && E_{ab}^{g}=\frac{1}{8\pi G_{N}} (R_{ab}-\frac12 g_{ab}R) -T_{ab}^{m}
.\nonumber \eea   

On the other hand, if we introduce  $\bw$, a  2 form in phase space but  codimension 1 form in spacetime, 
by 
\be
\bw (\phi;\delta_{1}\phi, \delta_{2}\phi)= \delta_{1}\Th(\delta_{2}\phi)-\delta_{2}\Th(\delta_{1}\phi) ,
\ee
we can express $\bfJ_\xi$  in terms of $\bw$ as follows 
\be 
\delta \bfJ_\xi= \bw(\d\phi,\d_{\xi}\phi) +d(\xi\cdot\Th(\delta\phi))-\xi\cdot\bfE^{\phi} \d\phi \label{dJ}
\ee
Using Eqs. (\ref{J2}) and (\ref{dJ}), we get an off-shell relation 
\bea 
 d \bchi &=&\bw(\d\phi,\d_{\xi}\phi) - \xi\cdot (\delta \bfC + \bfE^{\phi} \d\phi )\label{HW}, \cr
\hbox{ with } \quad \bchi &=& \delta \bfQ_\xi-\xi\cdot\Th(\delta\phi). 
\eea 
So far $\delta$ is infinitsimal variation defined by  $\delta\phi=\dla \phi(x;\lambda)|_{\lambda=0}$. 
The   point of  Holland and Wald  \cite{Hollands:2012sf}  is that if we replace $\delta \to \dla$ without setting $\lambda=0$ after derivative,  all the steps above 
go through so that we now have all order relation in $\lambda \in [0,\ve]$.  Then  Eq. \rf{HW} can be replaced by 
\bea 
 d \bchi &=&\bw(\dla\phi,\d_{\xi}\phi) - \xi\cdot (\dla \bfC + \bfE^{\phi} \cdot \dla\phi ), \hbox{ with } \cr
 \bchi &=& \dla \bfQ_\xi-\xi\cdot\Th(\dla \phi). \label{HW2}
\eea 
An important remark   is that we should work in Holland-Wald gauge \cite{Hollands:2012sf} where the Ryu-Takayanagi surface and    $\xi$ does not change its coordinate  dependence
 for any metric deformation $g(x;\lambda)$ with $\lambda\in [0,\ve]$,   which gives the restriction to the size of $\ve$.

Notice also that {for the linear order}  the canonical energy term becomes 
$\bw(g_0; \d g,\d_{\xi_{B}}g_0)$  and it vanishes for  AdS metric $g_0$ since $\d_{\xi_{B}}g_0=0$.  
Notice also 
in  Eq. (\ref{linearE}),  $\xi\cdot \bfE^{g}\cdot \d g$ does  not appear either,  because 
 the explicit form of AdS metric was already used to give  $\bfE[{g_0}]=0$.
However, for non-linear order, one has to consider a finite variation $g(\ve)$ and consider the cotangent space of  the space of metric at $g(\lambda)$  for arbitrary $\lambda$ between $0$ and $\ve$.   
In this case none of the two vanish and this fact  provides  the main source  of the non-triviality in getting non-linearity  of the gravity equation.


\section{Non-linear Einstein equation  from Entanglement}
The issue of full Einstein equation was   discussed earlier    in 
\cite{Swingle:2014uza,Jacobson:2015hqa,Dong:2017xht,Sarosi:2017rsq}  and most notably in \cite{Faulkner:2017tkh}, where the program of getting gravity equation starting from CFT 
 is extended perturbatively to second order. 
Essential part of above paper is to derive the gravity expression of relative entropy  starting from the CFT up to the second order. Similar efforts have been made in \cite{Sarosi:2017rsq}.
Given the fact that  completing this program to all order is certainly non-trivial, 
one may ask that if this part is assumed to be proven to all order, then 
   can  we  actually  show that the   full-non-linear   Einstein equation can be implied from there.  
This question can be addressed purely in gravitational context, because 
  as we will see shortly, the  gravity expression of relative entropy entropy can be derived from the Holland-Wald offshell identity by imposing the Einstein equation. One can ask the reverse question, namely, can we derive the Einstein equation from the 
relative entropy expression. We will see that the answer is positive. 

To simplify the setting  we consider only pure  gravity  so that $\phi(x;\lambda)$ is replaced by metric $g(x;\lambda)$, and choose $\xi$ as  the Killing vector of AdS given in Eq. (\ref{killing}).  
 
%
%

Integrating both side of Eq. \rf{HW} over $\Sigma$ whose boundary is $B$ and $\tilde B$, we get  Eq. (\ref{EE1st}) and (\ref{linearE}).  
By integrating \rf{HW2} over $\Sigma$, the region between $B$ and $\tilde B$ at time slice $t=0$,  we have   \cite{Hollands:2012sf,Lashkari:2016idm}
\def\veps{{\varepsilon}}
\bea 
 \int_{ B} {\boldsymbol \chi} -\! \! \int_{\tilde B} {\boldsymbol \chi} &=&
\! \! \int_{\Sigma}\bw(g_\lambda;\dla g_\lambda,\d_{\xi_{B}}g_\lambda) +\! \! \int_{\Sigma} ({\hat E}+{\hat C}), \label{HW3} \\
\hbox{where } {\hat E} &=&- {\xi^{a}_B}  {\beps}_{a} E^{g}_{bc}[g_\lambda]  \dla g^{bc}, \; {\hat C}= - {\xi^{a}_B}   \dla  \bfC_{a}[g_\lambda] .   \nonumber
\eea

First consider only  physical metrics which satisfy  equations of motion, then
 $\hat{ E}={\hat C} =0$ so that  
 \bea 
 \int_{ B} {\boldsymbol \chi} -  \int_{\tilde B} {\boldsymbol \chi}  = \int_{\Sigma}\bw(g_\lambda;\dla g_\lambda,\d_{\xi_{B}}g_\lambda)\label{HW33} .
\eea 
Notice that the right hand side is not zero since $ \xi_{B}$ is Killing vector of the background metric $g_0$
not that of $g_\lambda$. 
One should also notice that the first term of \rf{HW33} is not a total variation as one can see in \rf{HW2}
and therefore can not be written in general as $\dla E^{grav}_B$, while the second term is always a total variation so that it can be written as 
 $\dla S^{grav}_B$. 
 Integrating the Eq. \rf{HW33} by $\int_0^\ve d\lambda$,  we have 
\bea 
\Delta E^{grav}_{B} -\Delta S^{grav}_{B} = \int_0^\ve d\lambda \int_{\Sigma}\bw(g;\dla g,\d_{\xi_{B}}g) ,\label{HW4} 
\eea 
\vskip -.2cm
{where  }
\be 
\Delta E^{grav}_{B} = \int_0^\ve d\lambda  \int_{ B} {\boldsymbol \chi}, \quad
 \Delta S^{grav}_{ B} = \int_0^\ve d\lambda  \int_{ \tilde B} {\boldsymbol \chi}.\nonumber
 \ee
Since one can 'define'  the  relative entropy as the difference of $\Delta E$ and $\Delta S$ 
  as  we noted earlier, 
eq.(\ref{HW4}) can be used to identify  the gravity version of relative entropy \cite{Lashkari:2016idm},  
 \be
 S^{grav}(\rho |\rho_0)= \int_0^\ve d\lambda \int_{\Sigma}\bw(g;\dla g,\d_{\xi_{B}}g). \label{HW5} 
 \ee
Then,   Eq. \rf{HW4}  becomes    
\be 
\Delta E^{grav}_{B} -\Delta S^{grav}_{B} = S^{grav}(\rho |\rho_0), \label{HW66}
\ee
 which is nothing but  the gravity dual  of the generalized first law \rf{gen1st}.  
 \vskip.3cm 

 So far, we have seen  that the on-shell expression of Holland-Wald identity gives the gravitational version of the generalized first law. This has been known \cite{Faulkner:2017tkh,Lashkari:2016idm,Lashkari:2015hha,VanRaamsdonk:2016exw}.
 The authors of  \cite{Faulkner:2017tkh} 
     proved  differential version of \rf{HW4} from the CFT upto second order, which enabled them to prove the Einstein equation to the corresponding order. 

What we want to do is the reverse direction:  if a metric satisfies the gravity version of generalized 
entanglement first law, it should satisfy Einstein equation. 
In other words, we want to prove that the   gravity expression of the relative entropy, eq.(25) or its consequence (24),  is equivalent to the equation of motion.  
This is not a  totology.  Notice that deriving (25) from CFT is not our goal. 

Namely, we want to derive the  full Einstein equation, 
starting from Eq.(\ref{HW4}). This is {\bf our main goal}. 
  
By integrating   Eq.  \rf{HW3} in $\lambda$ over  $[0.\ve]$, we first rewrite it as 
\bea 
\Delta E^{grav}_{B} \!\!  -\!\! \Delta S^{grav}_{B} \!\! -  S^{grav}(\rho |\rho_0) 
=\!
\int_0^\ve\!\! d\lambda \!\!  \int_{\Sigma} ({\hat E} +{\hat{ C}}),     \label{HW88}
\eea 
%
Now if we impose  Eq. \rf{HW4} or \rf{HW66}, which is the gravity dual of the generalized first law of entanglement, the right hand side of above equation   vanishes. 
Taking the derivative of equation  with respect to $\ve$, we  get
 \be 
 {\hat E} [g(\ve)]  +{\hat{ C}}  [g(\ve)]=0.
 \ee
  Using the explicit form 
of the constraint given in \rf{constraint}, we have 
\be
\xi_b { E}^{cd}    [g(\ve)] g'(\ve)   + 2\xi^a E'_{ab}[g(\ve)]=0 . \label{shell}
 \ee 
where the prime denote $\frac{d}{d\ve}$ and we deleted the subscript/superscript $g,B$ from the $E$ to simplify the notation.   
We expand the $E_{ab}[g(\ve)]$ and $g_{ab}(\ve)$ in   $\ve$:  
\be
E[g(\ve)]=\sum_{n=0}^\infty \ve^n E^{(n)}, \quad \hbox{and } g(\ve)=\sum_{n=0}^\infty \ve^n g^{(n)}. 
\ee 
 Then   Eq.\rf{shell} becomes 
 \be
 \sum_{n=1}^\infty \ve^{n-1}\Big[ \xi_b \sum_{k=1}^{n} k E^{(n-k)}[g_0] \cdot g^{(k)} + 2\xi^a nE_{ab}^{(n)} \Big] =0 , 
  \ee
   where $\cdot$ is for the full contraction. 
 Requesting the analyticity in $\ve$, each coefficient of above equation should be zero. 
 It is useful to write the first few terms explicitly to see the structure: 
 \bea
 \xi_b   E^{(0)}[g_0]^{cd} g_{cd}^{(1)} + 2\xi^a E_{ab}^{(1)}=0, \cr
  \xi_b   (2E^{(0)}\cdot g^{(2)} +E^{(1)} \cdot  g^{(1)} ) + 4\xi^a E_{ab}^{(2)}=0, \cr
 \xi_b   (3E^{(0)} \cdot g^{(3)}+ 2E^{(1)} \cdot g^{(2)} +E^{(2)}  \cdot g^{(1)} )    + 6\xi^a E_{ab}^{(3)}=0,
\nonumber
 \eea
 \be
 \cdots .
 \ee
Notice that this is the expansion around the AdS metric $g_0$, so that 
 $E^{(0)}[g_0]=0$, which implies $E^{(1)}[g_0]=0$ by the first equation, which in turn implies $E^{(2)}[g_0]=0$ by the second  equation. In this way,  all     $E^{(n)}[g_0]=0$ by the lower ones progressively, proving the whole non-linear Einstein equation 
\be
\bfE[g(\varepsilon)]=0,\label{fE}
\ee 
for all order in $\varepsilon$. 
\footnote{It is worthwhile to notice that the same conclusion can be derived 
in more complicated situation where the eq. \rf{shell} is modified to 
\be 
 {\hat E} [g(\ve)] A(\ve) +{\hat{ C}}  [g(\ve)]B(\ve)=0,
 \ee
if the objects $A,B$ have expansion starting from $\ve^0$.
 }
Therefore, the metric $g(\varepsilon)$  near  $g_{0}$ satisfies full Einstein equation.
 
%

Summarizing,    the full Einstein equation holds iff the generalized entanglement first law does, thanks to  the  geometric  identity Eq. (\ref{HW88}). 
In other words,  the metric $g$  dual to the   state $\rho$ compatible with the generalized first law 
  satisfies  the   non-linear Einstein equation. 
  Although \rf{HW4} is derived using Einstein Equation, 
  it is special so that it can imply the Einstein equation itself through the  geometric identity. 
  What is the implication of all this? 
    It just means that the relative entropy (RE) expression or the generalized entanglement entropy contains on-shell information. 
    This is clear from the linearized level. There, first law implies on-shell condition. 
    The same should be true here. 
In fact,  in CFT side, the RE can be evaluated only for   physical configuration.   
Therefore on-shell information is hidden in the entanglement relationship. 
From gravity side, the Einstein equation is the criterion to judge whether a given metric configuration is physical.  
Therefore it is not surprising   expression of RE  encodes the information of on-shell-ness.

 \vskip .2cm
 One important remark is that while $ {\boldsymbol \chi}$ is a total derivative  $\lambda$ on  ${\tilde B}$ due to the vanishing of $\xi$ on ${\tilde B}$,  it  is not so on $B$.  Therefore $\Delta S^{grav}$ is a total variation but $\Delta E^{grav}$ is not so in general. 
This is exactly the same property of $\Delta E , \Delta S$ in CFT side as we emphasized earlier. 
However, for an integrable case where  $\int_{\Sigma} \xi \cdot\bw=0$, the situation is better, because  there exist   $K$ and $W_\xi$  such that 
$\xi_B \cdot \Theta(\dla g)=\dla(\xi_B\cdot K)$  and 
$ W_\xi = \int_{B-{\tilde B}}  (\bfQ-\xi\cdot K)$  respectively  \cite{Wald:1999wa}, so that 
  we can rewrite \rf{HW33} as  \cite{Lashkari:2015hha,Jafferis:2015del, Lashkari:2016idm}
\bea
 \dla W_\xi =  \int_{\Sigma}\bw(g;\dla g,\d_{\xi_{B}}g). \label{integrable}
 \eea 
This  can be integrated over $\lambda$ to give 
\bea 
  \Delta\! E^{grav}_{B} - \Delta\! S^{grav}_{B}=\Delta W_\xi  . \label{HW6} 
\eea 
where $\Delta$ is a variation from $\rho_0$ to $\rho$ whose dual geometries   are   
$g_{0}, g$ respectively.    This means that,  for an integrable case,    
 the relative entropy is a total variation and it can be interpreted as  the work done on the system to change it from $\rho_0$  to  $\rho$.  

Our method can be easily generalized to the case with inclusion of matter or higher derivatives. 
For  the reference states other than the AdS vacuum, 
the  barrier  is  the proof of the existence of the Killing vector and its Holland-Wald gauge condition.  
We leave these matter to the  future works. 

\vskip.2cm
 
\section{Entanglement Vector Field}
In ref. \cite{Freedman:2016zud}, the authors tried to reformulate 
entanglement   as the a flux of vector field $\bf v$. 
Consider a surface $B'$ in $t=0$ slice whose boundary is the same as that of $B$. 
Our goal is to  construct a vector field  $V_E$ such that 
\be
\int_{B'}V_E^{a} d{\sigma_{a}} = \int_{\tilde B}V_E^{a} d{\sigma_{a}}=S_{B}. \label{div}
\ee 
Such vector field should  be divergenceless in the subspace of $t=0$ slice. 
Also it must be a codimension 2 form to  produce an one form upon restriction.
Natural candidate is   $*\bfQ$  restricted to the constant time slice and we start from the observation   
\be 
\int_{{\tilde B}} \bfQ =S_{B}, \quad d\bfQ  = -\xi\cdot \beps L \neq 0, 
\ee
 on shell,   where  we used   Eq.(\ref{J2}) 
 and  the  fact that   $\xi$ is the Killing vector of $g$.   
%
Now we can construct a vector field $V$ by restricting
 the codimension 2 form $\bfQ$ to the $t=0$ slice. Noticing that among the components of $\xi$, only $\xi^{t}$ is non-zero,  we have 
\bea
16\pi G_{N}{\bf Q}=\nabla^a\xi^b \e_{ab}  =-2 \nabla_{a}\xi^{t}  \sqrt{-g_{tt}}{  \epsilon}^{a} := V_{a}\epsilon^{a} .
\eea 
In one form notation, the $V_{a}$ is give by 
\be
V= \frac{4\pi}{Rz}\Big[(\frac{R^{2}-z^{2}-{\vec x}^{2}} {2z}+z)dz+x^{i} dx^{i} \Big].
\ee
It is easy to check that $\int_{\tilde B} V_{a}\epsilon^{a} =4\pi {\rm Area}[\tilde B]$.
Therefore it is tempting to call $V_{a}$ as entanglement  vector field. 
However, for a vector field to be interpreted as a flux, it should be  
  divergenceless so that the flux on arbitrary surface $B'$ is equal to $S_{B}$. 
  Unfortunately, $V$  is not divergence free.  In fact, in $t=0$ slice of AdS$_{d+1}$,
\be
\nabla_{a}V^{a}=\frac{2\pi d}{Rz}   (z^{2}+{\vec x}^{2} -R^{2} )=(-2d) n\cdot \xi,
\ee
where $n$ is the normal vector of the hypersurface $\Sigma$. Furthermore, 
while we expect that the entanglement vector's  flux is highly  concentrated at the  boundary of 
the region $B$,   the  flux of $V$, as one can see in the Fig.1,  is almost uniformly distributed over $B$. 
\begin{figure}[ht!]
\centering
    \subfigure[ ]
   {\includegraphics[width=3.2cm]{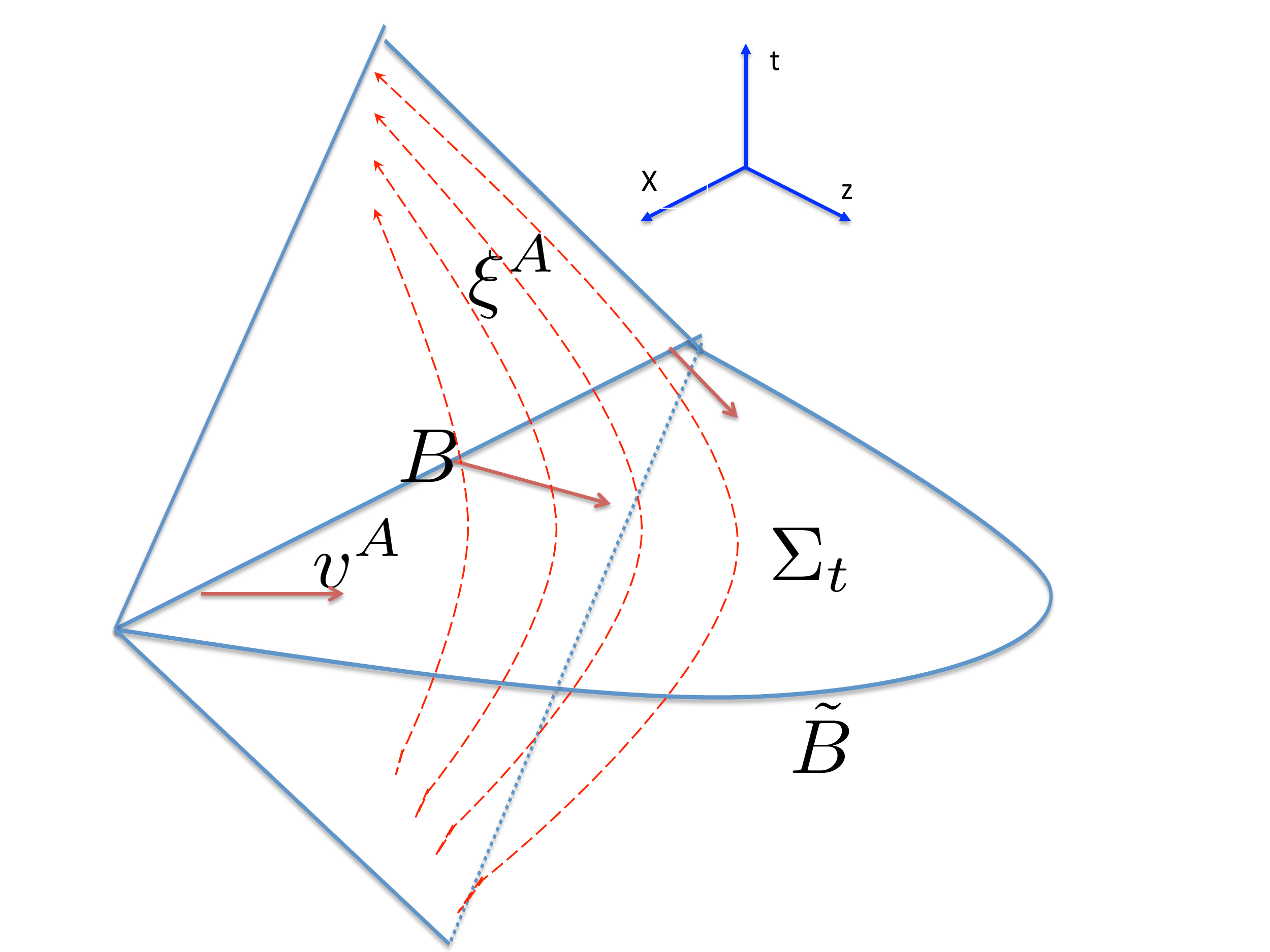} \label{}}     \hspace{.1cm}
    \subfigure[ ]
   {\includegraphics[width=4.8cm]{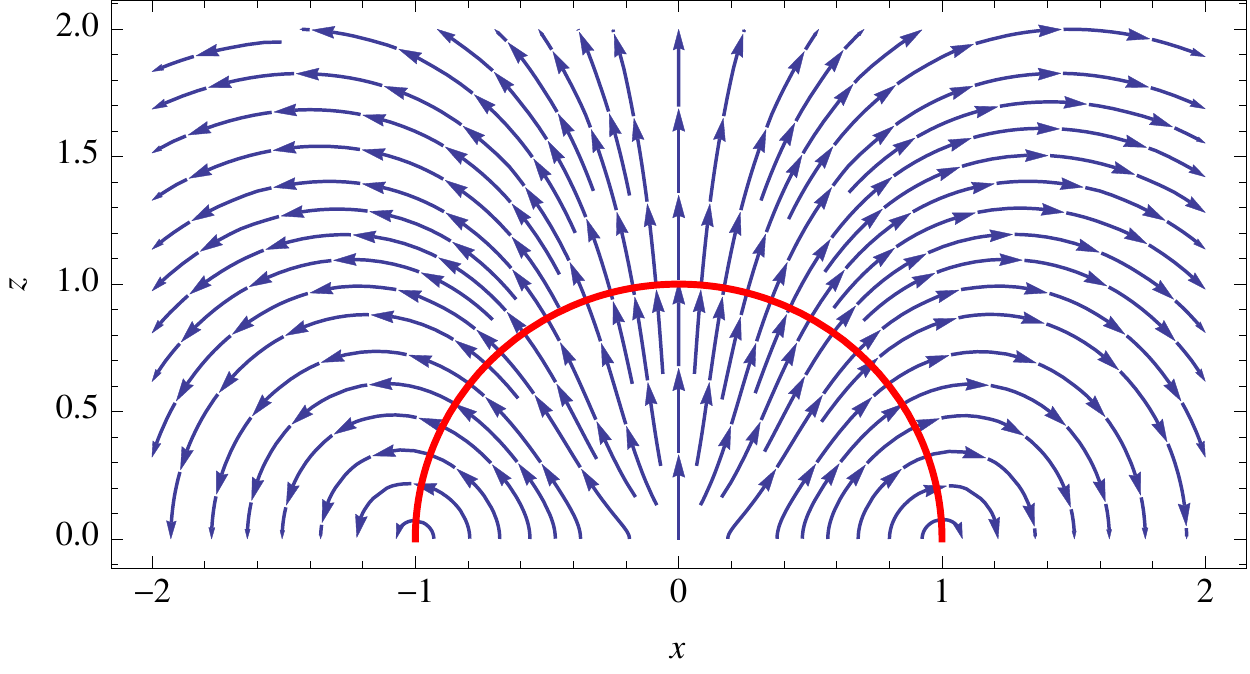} \label{}}     \hspace{.1cm}
     \caption{  (a)Entanglement wedge and  flow of  vector field   $\xi$ and $V$. 
     (b)Flow of  vector field   $V$ within $\Sigma$. The red circle is the Ryu-Takayanagi surface } \label{fig:V}
\end{figure}  

Therefore we look for a balancing vector field $V_{0}$ such that $\nabla_{a}(V^{a}-V_{0}^{a}) = 0$ and 
  flux of $V_{0}$ over $\tilde B$ is zero. We take ansatz  $V_{0}=V_{0r}dr$  and  boundary condition  
  $V_{0r}\big |_{r=R}=0$.  One remark  is that when we take the divergence of $V$, we should consider $\xi^{t}$ as a scalar once we restrict $\bfQ$  to $t=0$ slice.  
In $AdS_{d+1}$,  it can be  given by
\be 
V_{0}=\frac{2\pi d}{R} \frac{(r-R)^{2}}{r^{2}\cos^{3}\theta}   dr, 
\ee
where $r^{2}=z^{2}+{\vec x}^2$ and $\cos\theta=z/r$. 
The final form of the entanglement vector field  is give by $V_{E}=V-V_{0}$ whose explicit form in polar coordinate is 
\be
{ V_{E}}= \frac{2\pi}{R} \Big[  \frac{r^{2}+R^{2}}{r^{2}\cos\theta}dr -\frac{(R^{2}-r^{2})}{r}
\frac{\tan \theta}{\cos \theta} d\theta \Big] -V_{0}
\ee   
which is divergence free vector field whose flux over any $B'$ is $S_{B}$ if 
$B'$ is homologous to $B$.   
One can easily verify that $V_{E}$ satisfies Eq.(\ref{div}), and 
 for $AdS_{3}$ the flux of each vector fields are 
\be
\int_{B}V_{a}\beps^{a}= \frac{c}{9}\frac{R^{2}}{\epsilon^{2}}, \quad    
\int_{B}V_{0a}\beps^{a} =\frac{c}{9}\frac{R^{2}}{\epsilon^{2}}   -\frac{c}{3}\ln \frac{2R}{\epsilon},  \quad
 \ee
where  $\epsilon$ is the UV cut-off of $z$ and $c=\frac{3L}{2G_{N}}$ with $L$  the AdS radius and $c$ 
is the central charge of the dual  $CFT_{2}$. 
 \begin{figure}[ht!]
\centering
        \subfigure[ ]
 {\includegraphics[width=4.8cm]{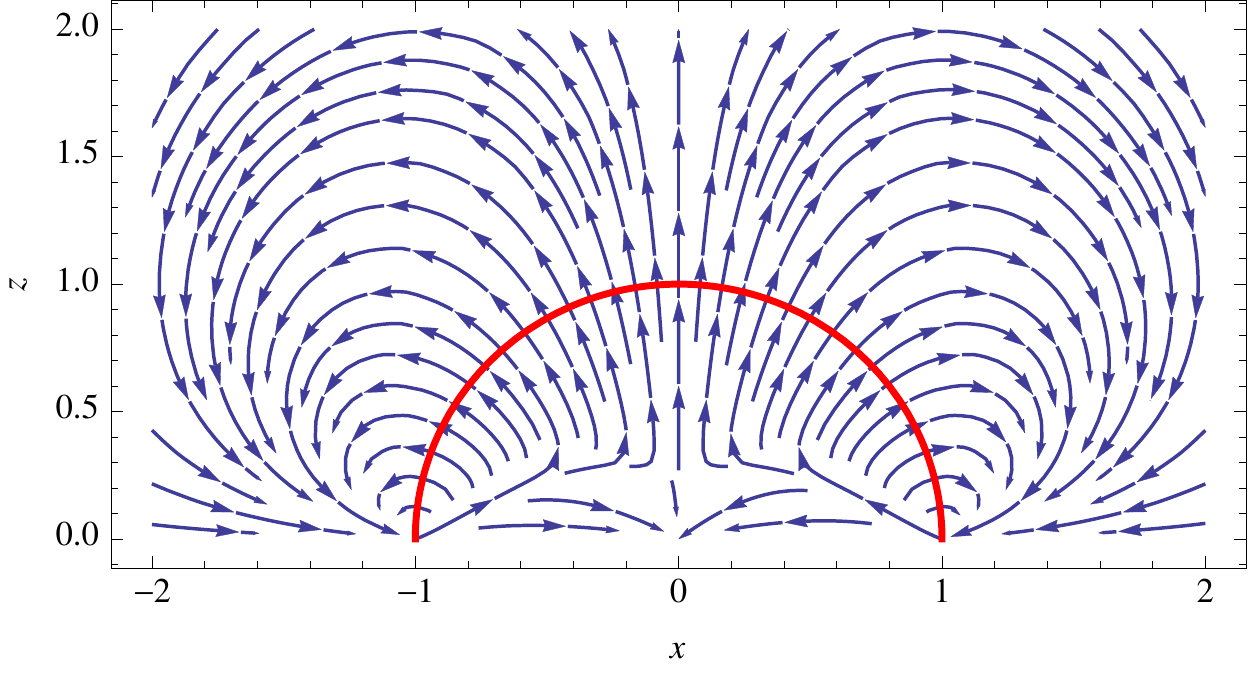} \label{}}
      \subfigure[ ]
 {\includegraphics[width=3.5cm]{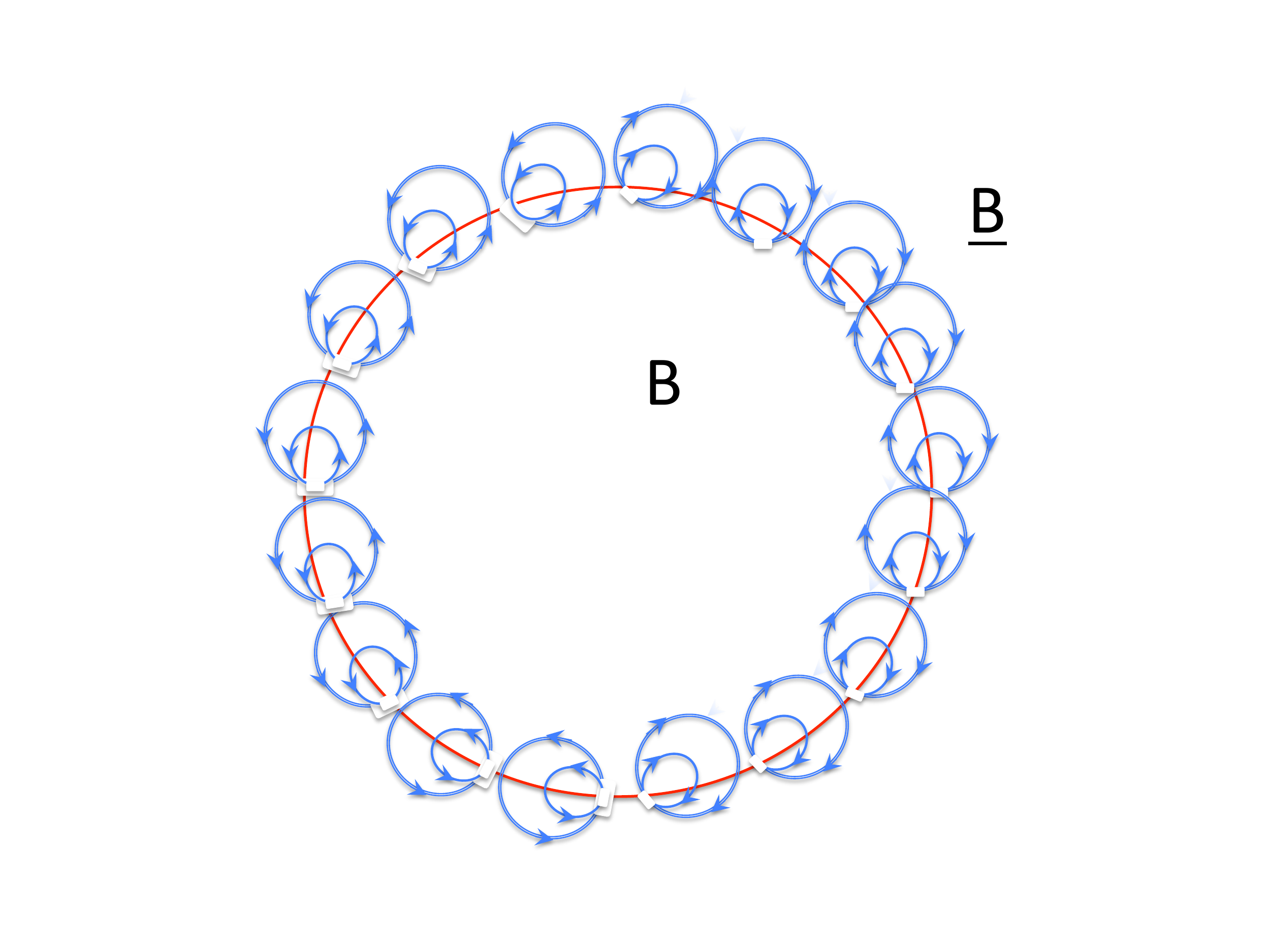} \label{}}
 \caption{  (a) Flow of  the Entanglement   vector  field $ V_{E}$. 
 (b) Cartoon   of 3d version of left figure where it  is rotated around z-axis.} \label{fig:VEE}
\end{figure}  

Our goal here is to explicitly construct  the thread vector of ref. \cite{Freedman:2016zud}, where the authors 
suggested to replace minimal surface by a divergenceless vector. Notice however,  
 the flux line in fig. 15 of ref.[21] is similar to our vector field $V$  in FIG. 1 which is  {\it not divergenceless.} 
  If we impose zero divergence condition, the resulting vector field $V_{E}$ has the flux lines    concentrated at the boundary of the two regions, which reveal  quite interesting phenomena: 
  entanglement is done mostly at the boundary of the two entangled regions.   
As a consequence, the flux of $ V_{E}$, as one can see in the Fig.2, look like sewing the two regions $B$ and $\bar B$ along their interface through the holographic direction, which is an anticipated feature for the entanglement entropy vector field but was not expected from the general argument of ref. \cite{Freedman:2016zud}.  


\section{ Discussion}
We   have shown that the generalized entanglement first law implies the full Einstein equation. It would be interesting to study  the case in  the presence of matter fields or higher curvature term.  
We also  constructed a vector field $V$ in $AdS$ space whose flux on arbitrary surface homologous to $B$ is equal to the  entanglement entropy.  It would be   interesting if we can utilize the entanglement vector flow to discuss the black hole information problem. 

 \vskip 0.1cm 
\ni {\bf Acknowledgments} 
SJS want to thank  Kimyeong Lee, Sungjae Lee and Piljin Yi
for discussion. This  work is supported by Mid-career Researcher Program through the National Research Foundation of Korea grant No. NRF-2016R1A2B3007687.  We  appreciate the hospitality of APCTP during the focus workshop   ``Geometry and Holography of Quantum critical point''.


\begin{thebibliography}{99}

\bibitem{VanRaamsdonk:2010pw} 
  M.~Van Raamsdonk,
  ``Building up spacetime with quantum entanglement,''
  Gen.\ Rel.\ Grav.\  {\bf 42}, 2323 (2010)
  [Int.\ J.\ Mod.\ Phys.\ D {\bf 19}, 2429 (2010)]
  [arXiv:1005.3035]


\bibitem{Maldacena:2013xja} 
  J.~Maldacena and L.~Susskind,
  ``Cool horizons for entangled black holes,''
  Fortsch.\ Phys.\  {\bf 61}, 781 (2013)
  [arXiv:1306.0533 [hep-th]].



\bibitem{Lashkari:2013koa} 
N.~Lashkari, M.~B.~McDermott and M.~Van Raamsdonk,
``Gravitational dynamics from entanglement 'thermodynamics',''
JHEP {\bf 1404}, 195 (2014)
doi:10.1007/JHEP04(2014)195
[arXiv:1308.3716]

\bibitem{Ryu:2006ef} 
S.~Ryu and T.~Takayanagi,
``Aspects of Holographic Entanglement Entropy,''
JHEP {\bf 0608}, 045 (2006)
[hep-th/0605073].

 
\bibitem{Casini:2011kv} 
H.~Casini, M.~Huerta and R.~C.~Myers,
``Towards a derivation of holographic entanglement entropy,''
JHEP {\bf 1105}, 036 (2011)
[arXiv:1102.0440 [hep-th]].



\bibitem{Faulkner:2013ica} 
T.~Faulkner, M.~Guica, T.~Hartman, R.~C.~Myers and M.~Van Raamsdonk,
``Gravitation from Entanglement in Holographic CFTs,''
JHEP {\bf 1403}, 051 (2014)
[arXiv:1312.7856 [hep-th]].

\bibitem{Wald:1993nt} 
R.~M.~Wald,
``Black hole entropy is the Noether charge,''
Phys.\ Rev.\ D {\bf 48}, no. 8, R3427 (1993)
doi:10.1103/PhysRevD.48.R3427
[gr-qc/9307038].



\bibitem{Iyer:1994ys} 
V.~Iyer and R.~M.~Wald,
``Some properties of Noether charge and a proposal for dynamical black hole entropy,''
Phys.\ Rev.\ D {\bf 50}, 846 (1994)
doi:10.1103/PhysRevD.50.846
[gr-qc/9403028].

\bibitem{Iyer:1995kg} 
V.~Iyer and R.~M.~Wald,
``A Comparison of Noether charge and Euclidean methods for computing the entropy of stationary black holes,''
Phys.\ Rev.\ D {\bf 52}, 4430 (1995)
doi:10.1103/PhysRevD.52.4430
[gr-qc/9503052].

\bibitem{Wald:1999wa} 
  R.~M.~Wald and A.~Zoupas,
  ``A General definition of 'conserved quantities' in general relativity and other theories of gravity,''
  Phys.\ Rev.\ D {\bf 61}, 084027 (2000)
  doi:10.1103/PhysRevD.61.084027
  [gr-qc/9911095].
  
\bibitem{Hollands:2012sf} 
  S.~Hollands and R.~M.~Wald,
  ``Stability of Black Holes and Black Branes,''
  Commun.\ Math.\ Phys.\  {\bf 321}, 629 (2013)
  doi:10.1007/s00220-012-1638-1
  [arXiv:1201.0463]
 
\bibitem{Jacobson:1995ab} 
  T.~Jacobson,
  Phys.\ Rev.\ Lett.\  {\bf 75}, 1260 (1995)
  doi:10.1103/PhysRevLett.75.1260
  [gr-qc/9504004].
  

\bibitem{Faulkner:2017tkh} 
  T.~Faulkner, F.~M.~Haehl, E.~Hijano, O.~Parrikar, C.~Rabideau and M.~Van Raamsdonk,
  ``Nonlinear Gravity from Entanglement in Conformal Field Theories,''
  JHEP {\bf 1708}, 057 (2017)
  doi:10.1007/JHEP08(2017)057
  [arXiv:1705.03026 [hep-th]].
 


\bibitem{Blanco:2013joa} 
D.~D.~Blanco, H.~Casini, L.~Y.~Hung and R.~C.~Myers,
``Relative Entropy and Holography,''
JHEP {\bf 1308}, 060 (2013)
[arXiv:1305.3182] 

\bibitem{Lin:2014hva} 
  J.~Lin, M.~Marcolli, H.~Ooguri and B.~Stoica,
  ``Locality of Gravitational Systems from Entanglement of Conformal Field Theories,''
  Phys.\ Rev.\ Lett.\  {\bf 114}, 221601 (2015)
  [arXiv:1412.1879 ]


\bibitem{Lashkari:2016idm} 
  N.~Lashkari, J.~Lin, H.~Ooguri, B.~Stoica and M.~Van Raamsdonk,
  ``Gravitational positive energy theorems from information inequalities,''
  PTEP {\bf 2016}, no. 12, 12C109 (2016)
  [arXiv:1605.01075 [hep-th]].
  
\bibitem{Lashkari:2015hha} 
  N.~Lashkari and M.~Van Raamsdonk,
  ``Canonical Energy is Quantum Fisher Information,''
  JHEP {\bf 1604}, 153 (2016)
  doi:10.1007/JHEP04(2016)153[arXiv:1508.00897 ]
  
  \bibitem{VanRaamsdonk:2016exw} 
  M.~Van Raamsdonk,
  ``Lectures on Gravity and Entanglement,''
  doi:10.1142/9789813149441-0005
  arXiv:1609.00026 [hep-th].
  
 \bibitem{Jafferis:2015del} 
  D.~L.~Jafferis, A.~Lewkowycz, J.~Maldacena and S.~J.~Suh,
  JHEP {\bf 1606}, 004 (2016)
  doi:10.1007/JHEP06(2016)004
  [arXiv:1512.06431 [hep-th]].
  
  

 

\bibitem{Swingle:2014uza} 
  B.~Swingle and M.~Van Raamsdonk,
  ``Universality of Gravity from Entanglement,''
  arXiv:1405.2933 [hep-th].
  
\bibitem{Jacobson:2015hqa} 
  T.~Jacobson,
  ``Entanglement Equilibrium and the Einstein Equation,''
  Phys.\ Rev.\ Lett.\  {\bf 116}, no. 20, 201101 (2016)
  doi:10.1103/PhysRevLett.116.201101
  [arXiv:1505.04753 [gr-qc]].
  
  \bibitem{Sarosi:2017rsq} 
  G.~Sárosi and T.~Ugajin,
  ``Modular Hamiltonians of excited states, OPE blocks and emergent bulk fields,''
  arXiv:1705.01486 [hep-th].
 
 
\bibitem{Dong:2017xht} 
  X.~Dong and A.~Lewkowycz,
  ``Entropy, Extremality, Euclidean Variations, and the Equations of Motion,''
  arXiv:1705.08453 [hep-th].
  


\bibitem{Freedman:2016zud} 
  M.~Freedman and M.~Headrick,
  ``Bit threads and holographic entanglement,''
  arXiv:1604.00354 [hep-th].


\end{thebibliography}
\end{document}